\documentclass[showpacs,aps,prl,preprint,floatfix]{revtex4}
\usepackage{amsfonts}
\usepackage{amssymb}
\usepackage[fleqn]{amsmath}
\usepackage[mathscr]{eucal}
\usepackage{graphicx}

\newcommand{\vect}[1] {\boldsymbol{{ #1}} }
\newcommand{\pVN}{{\vect{P}}}
\newcommand{\xVN}{{\vect{X}}}
\newcommand{\rVN}{{\vect{R}}}
\newcommand{\pV}{{\vect{p}}}
\newcommand{\xV}{{\vect{x}}}
            
\newcommand{\dij}{\xV_i-\xV_j}

\newcommand{\dik}{\xV_i-\rVN_k}

\begin{document}

\title{Field-induced phases of an orientable
charged particle \\ in a dilute background of point charges.}

\author{Carlo Lancellotti}
\affiliation{Graduate Faculty in Physics \& Department of Mathematics, City University of New York-CSI, Staten Island NY 10314}
\author{Bala Sundaram}
\affiliation{Department of Physics, University of Massachusetts, Boston, MA 02125}

\date{\today}

\begin{abstract}
We study a dynamical model of a rod-like particle surrounded by a cloud of 
smaller particles of the same charge and we
show that, in the presence of a low-frequency alternating electric field, 
the rod displays the same type of anomalous orientation (perpendicular
to the field) that was recently observed in laboratory colloids. 
This indicates that the anomalous orientation is due to the collective 
dynamics of the colloidal particles, and does not require electro-osmotic 
effects.  We also confirm the experimental observation that for higher field 
frequencies the standard orientation (parallel to the field) prevails.  
In the simulations, these changes are abrupt enough to resemble a phase 
transition.
\end{abstract}

\pacs{45.50.-j, 82.70.Dd, 83.10.Rs}
\maketitle

An important feature of many colloidal particles is that they are electrically
charged, due to surface effects involving their interaction
with the solvent or with an electrolyte \cite{saville,hunter}.  
The long-range nature of electrostatic interactions generates
interesting collective or ``electrokinetic'' effects
that resemble in many ways the complex phenomena that are familiar
in plasma physics.  In particular, colloidal dynamics can be very sensitive
to applied external electric fields. For example, it is known \cite{kramer, cates}
that at high concentrations rodlike colloids display field-induced anomalous (negative) optical 
birefringence. This implies that the rods align perpendicular to the external field,
whereas  at lower densities they assume the more intuitive alignment parallel to
it.  This phenomenon of anomalous orientation has been the object of some 
theoretical studies \cite{cates,blair,melissa}, but remains basically unexplained. 
Its importance has been recently underscored by some new experimental work \cite{cecco}, 
which studied how \emph{dilute} suspensions of charged rod-like colloids 
(``primary particles," or PP) respond to a low external electric field in
the presence of smaller spherical charged particles (``secondary
particles," or SP).  Once again, the field-induced orientation of the rod-like 
colloids was investigated by measuring the optical birefringence of the
solution and extracting the Kerr constant as a function of
the frequency of the field.  The key result was that, when the SP
are present, even in dilute solutions the PP align perpendicular
to the field as long as the forcing frequency is lower that a certain
threshold.  This is a surprising result, especially because this ``anomalous" 
orientation seems to be universal \cite{cecco} in mixtures of this 
type, suggesting that there is a general physical mechanism in need of 
theoretical explanation.  

Here we make the case that this phenomenon does reflect
a basic and universal dynamical effect, by showing how
the anomalous orientation arises already in a very simple two-dimensional
model, in which a single one-dimensional rod with charges at both 
ends interacts with a cloud of point charges. All the charges in question
are taken to have the same sign; the whole system is driven by an alternating
electric field and is placed in a box with periodic boundary conditions.
Obviously, such a simplified model cannot be expected to yield quantitatively
accurate predictions of the behavior of laboratory colloids.  On the other
hand, it is precisely the simplicity of the model that makes it 
significant that we obtain the anomalous orientation
that has been observed experimentally \cite{cecco}.
This suggests that the effect under consideration is quite fundamental
and independent of the detailed structure of the colloids. In fact,
the sophisticated  electro-osmotic phenomena \cite{cecco2} that take place around real 
colloids are completely absent from our model. Hence, these effects are shown not to 
play an essential role in changing the orientation, since the collective dynamics 
alone of the rod and the particles are able to do it.  Our simulations
suggest the following scenario: due to the relative
motion induced by the field, when the bar is not perpendicular 
to the field the charges at its two ends compress and decompress 
asymmetrically the cloud of SP.  Such asymmetry in the density of the SP
generates a collective torque that tends to push the 
bar toward the perpendicular alignment until the symmetry is restored.  
This mechanism, however, is effective only if the frequency of the field 
is low enough, because if the bar oscillates too quickly the 
SP cannot organize collectively in a torque-producing configuration, and the system
enters a regime in which the bar adopts the more familiar
orientation along the field.  Interestingly, the transition
from the ''anomalous" to the "regular" orientation is quite abrupt, so 
much so that it is reminiscent of a first-order phase transition.
The orientation is fairly 
independent of the polarizability of the bar and the amplitude of the field.
However, the aspect ratio of the container appears to be important.

\section{Dynamical Model}
\vspace{-0.15in}
We introduce a two-dimensional molecular-dynamics-type model of a colloidal 
suspension containing a rod-like colloid and multiple (identical) point 
particles.  The Hamiltonian for this simplified model is
\begin{eqnarray}
&& H(\xVN,\pVN,\theta,p_\theta,\xV_1,\dots,\xV_N,\pV_1,\dots,\pV_N)=
\nonumber\\[0.2cm]
&&\frac{\left|\pVN\right|^2}{2M}+\frac{p_\theta^2}{2ML^2}+
\sum_{i=1}^n\frac{|\pV_i|^2}{2m}+
\nonumber\\[0.2cm]
&&\frac{1}{2}\sum_{i=1}^n\sum_{j\neq i}
\frac{q^2}{|\xV_i-\xV_j|}\,+
\sum_{k=1}^{2}\sum_{i=1}^n
\frac{q\,Q_k}{|\xV_i-\rVN_k|}\,+
\nonumber\\[0.2cm]
&&\left[\sum_{i=1}^n q\xV_i+\sum_{k=1}^{2}Q_k\rVN_k\right]\cdot\vect{j}\,
F\cos{\Omega t}
\end{eqnarray}
where $(\xVN,\pVN)$ are the canonical coordinates of the center of
mass of the bar, $(\xV_i,\pV_i)$ are the coordinates of the $i$-th secondary
particle for $i=1,\dots,n$ and  $\theta$ is the angle between the bar and the
$x$-axis; $\rVN_1\equiv\xVN+L\boldsymbol{\nu}$ and $\rVN_2\equiv
\xVN-L\boldsymbol{\nu}$ 
with $\boldsymbol{\nu}\equiv(\cos{\theta},\sin{\theta})$
are the positions of the ends of the bar, which has length $2L$; 
$\vect{j}$ is the unit vector in the direction of the external field.
The mass and charge of each SP are $m$ and $q$; the bar carries two masses
$M/2$ and two charges $Q_1$, $Q_2$ concentrated at each end.  
$F$, $\Omega$ are the amplitude and frequency of the forcing field.  

We introduce dimensionless variables by measuring
space, time, masses and charges in units of $L$, $\Omega^{-1}$, $m$ and
$q$, respectively.  The equations of motion become 
\setlength{\arraycolsep}{0.3mm}
\begin{eqnarray}
\ddot\xVN&=&-\frac{\alpha}{N}\sum_{k=1}^2\sum_{i=1}^n
\left[\tilde Q_k\frac{\dik}{|\dik|^3}\right]
-\frac{f}{N}\sum_{k=1}^2\tilde Q_k\,\cos t\,\vect{j}\nonumber\\[0.2cm]
\ddot\xV_i&=&
\alpha\!\sum_{j\neq i}\frac{\dij}{|\dij|^3}+
\alpha\!\sum_{k=1}^2\tilde Q_k\frac{\dik}{|\dik|^3}
-f\cos t\,\vect{j}\nonumber\\[0.2cm]
\ddot\theta&=&\frac{\alpha}{N}\sum_{k=1}^2\sum_{i=1}^n
\left[(-1)^k\tilde Q_k\frac{\xV_i-\xVN}{|\dik|^3}\cdot
\frac{d\boldsymbol{\nu}}{d\theta}
\right]\nonumber\\[0.2cm]
&-&\frac{f}{N}(\tilde Q_1-\tilde Q_2)\,\cos t\,\cos\theta
\label{NEWTeqs}
\end{eqnarray}
where $\tilde Q_i\equiv Q_i/q$, $N=M/m$ and 
\begin{equation}
\alpha\equiv\frac{q^2}{m\Omega^2 L^3}\qquad
f\equiv\frac{qF}{m\Omega^2 L}.
\end{equation}
Clearly, $\alpha$ is the ratio squared of the period of oscillation
of the field divided by the time scale over which the electrostatic repulsion 
between two SP's is able to move them across the length of the bar.  Hence,
$\alpha$ measures the coupling between secondary
particles.  As for $f$, it is the ratio squared of the period of oscillation
of the field divided by the time scale over which the field itself moves
a SP across the length of the bar; thus $f$ is a dimensionless measure of
the field strength.  Without loss of generality, we choose 
$m=q=L=1$, so that $\alpha=\Omega^{-2}$,
$f=F\Omega^{-2}$. 

One can add some simple enhancements to this model that make it
somewhat more realistic.  First of all, we will take into account
the effects of polarization by replacing the fixed charges $\tilde Q_k$, $k=1,2$,
in Eqs.\ (\ref{NEWTeqs}) with the functions
\begin{equation}
\tilde Q_k(\theta,t)=Q\,\big[1+(-1)^k \epsilon(\theta) F\cos t\big]
\end{equation}
where $0\leq \epsilon(\theta)\leq 1$ is an angle-dependent polarizability coefficient.
Also, we will add to the right-hand 
sides of Eqs.\ (\ref{NEWTeqs})
three Langevin terms $-\gamma_1\dot\xVN$, 
$-\gamma_2\dot\xV_i$ and $-\gamma_3\dot\theta$ in order to have
a crude model of the frictional effects of the solvent.  
One could also simulate the screening effect of an electrolyte by
replacing the Coulomb potential with a Yukawa potential 
$\phi(r)=e^{-\kappa r}/r$ where $\kappa$ is the inverse Debye length \cite{saville}.
Here, however, we will consider exclusively the Coulomb case $\kappa =0$ (no electrolyte),
modified only by a short-range cut-off for both physical reasons and numerical convenience.

In practice, the life-span of numerical solutions to Eqs.\ (\ref{NEWTeqs}) 
is seriously limited by the fact that the SP's impart a
slow (for $N\gg 1$) net drift to the center of mass of the bar.  When the bar
hits the box's wall, calculations with periodic boundary conditions
get disrupted because the bar gets ``broken," and one endpoint moved 
to the opposite side of the box.  Reflecting boundary conditions, on the 
other hand, interfere heavily with the rotation of the bar and make it hard
to observe the influence of the SP's and of the external field.
For the sake of simplicity, it is convenient to assume that the motion of 
$\xVN$ is determined only by the field and not by the SP's; this allows us 
to focus on
the crucial coupling between the \emph{rotational} degree of freedom of
the bar and the secondary particles.  If we drop the first term on the
right-hand side in the equation for $\xVN(t)$, we can choose the solution
\begin{equation}
\xVN(t)=\left(X_0,Y_0+\frac{2Qf}{N}\cos t\right)
\label{CDM}
\end{equation}
where $(X_0,Y_0)$ is the center of the box, 
substitute it into the other
equations in (\ref{NEWTeqs}) and solve them numerically.

\section{Numerical simulations}
\vspace{-0.15in}
\begin{figure}
\begin{center}
\includegraphics[height=4in,width=3.2in]{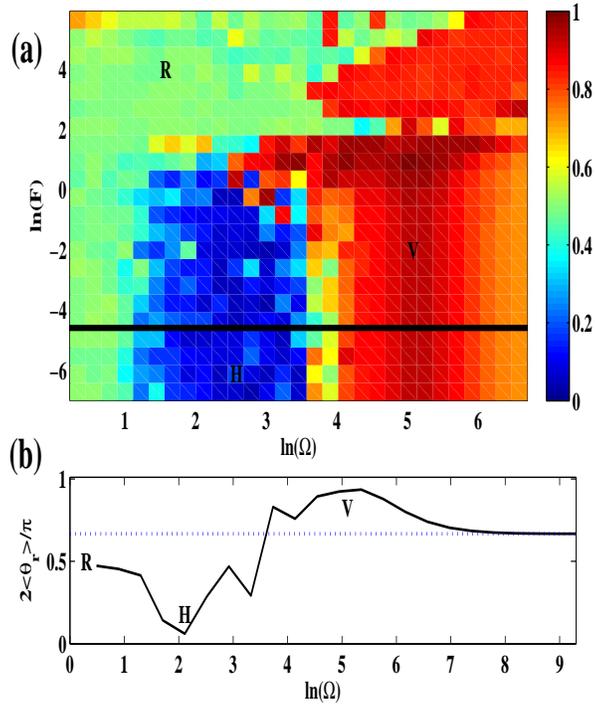}
\end{center}
\caption{\label{fig1} (a) Angular deviation of the bar from horizontal 
(measured by the reference angle $\theta_r$ in units of $\pi/2$, and 
time-averaged over the final ten cycles) shown as a function of external field 
parameters (amplitude $F$ and frequency $\Omega$). (b) Same deviation
plotted as a function of $\Omega$ with fixed $F=0.01$. The dotted line 
marks the orientation of the bar at $t=0$.}
\end{figure}

The equations we just introduced were solved numerically for $n=50$
SP's in a box with periodic boundary conditions.  Typical parameters
that roughly reflect the physical charge and mass ratios are obtained by choosing
$Q=N=10$.  Since our numerical experiments show that the preferred orientation
of the bar is not very sensitive to changes in either the frictional effects
or the polarizability of the bar itself, we also fix 
$\gamma_1=\gamma_2=\gamma_3=0.05$ and set $\epsilon (\theta)\equiv 0.2$ 
(neglecting the angular dependence of the polarizability). 
Thus, we are left with the two parameters $\alpha$ and $f$ -- or, 
equivalently, $\Omega$ and $F$.

The overall dependence of the bar's orientation on $\Omega$ and $F$ is shown
in Fig.~(\ref{fig1}a).  We characterize the orientation via the reference angle $\theta_r$ 
(the angle $\theta$ mapped onto the first quadrant), which is intuitively
more transparent than he usual $P_2(\theta)$ Legendre polynomial;   
$\big<\theta_r\big>$ denotes the average of $\theta_r$  over time.
In Fig.~(\ref{fig1}), the color red marks the points in the $\Omega$-$F$ plane
where the bar aligns along the direction of the field 
($\,\big<\theta_r\big>=\pi/2$, $P_2=-0.5\,$), whereas blue indicates
the parameter values that lead to anomalous orientation orthogonal to the field 
($\,\big<\theta_r\big>=0$, $P_2=1\,$).
The green points indicate that the time-averaged deviation from the horizontal
position is $\pi/4$ ($P_2=0.25$), which is the same value that one would get if the 
orientation
of the bar were just a uniformly distributed random variable.  Interestingly, the
parameter plane is divided in three well-defined regions, with fairly sharp boundaries,
where each one of these three behaviors (regular, anomalous and random orientation) 
is prevalent. For $\ln{F}\lesssim 0$ the choice of orientation is essentially
independent of $F$ and depends only on $\Omega$. At higher frequencies the bar
aligns with the external field, but if $\ln{\Omega}\lesssim 4$ the preferred orientation changes
to orthogonal to the field. At even lower frequencies, however, the external field
is not able to orient the bar at all and the angle appears to change randomly. Note that
increasing the frequency further ($\ln{\Omega}\gtrsim 6$) results in the bar moving away from a clear
vertical orientation (orange region). Dynamically speaking, this
is the high-frequency regime where the field polarity changes too rapidly and the
bar is minimally affected by both the field and the SP's. In this regime, as clearly seen from
Fig.~(\ref{fig1}b), the bar remains fixed at the initial orientation. Finally, for $\ln{F}\gtrsim 1$
only the ``regular"and "random" cases arise.
We also tested that in the absence of any SP
one gets only the vertical orientation (red), as expected.

\begin{figure*}
\includegraphics[height=6.5in,width=3.5in,angle=-90]{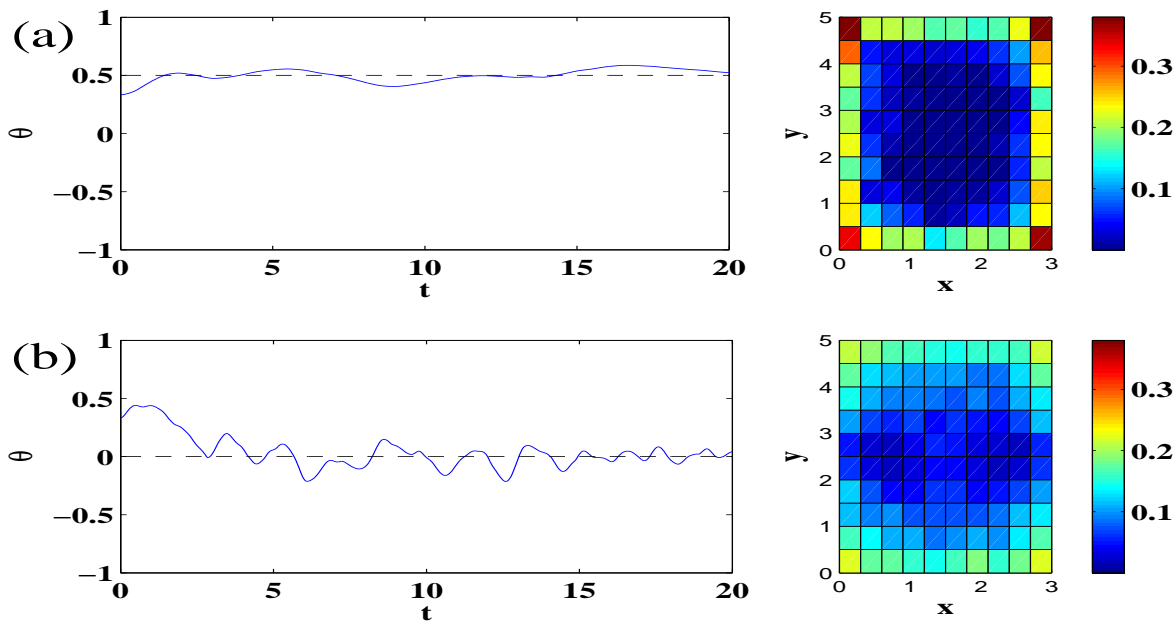}
\caption{\label{fig2} Contrast cases of (a) normal, $F=0.35,\Omega=7.5$ and (b) anomalous, 
$F=2.72, \Omega=33$,  orientation.  In each case, the dynamics of the orientation over twenty 
cycles of the field as well as the density distribution of the secondary (small) particles 
averaged over a cycle are shown. Note the distinct difference in distributions in the two 
situations. The dotted lines are intended to guide the eye in terms of final 
orientation and the angle is measured in units of $\pi$.}
\end{figure*}

In order to gain some insight into the particle dynamics associated with
the regular and the anomalous orientation, in Fig.~(\ref{fig2}) we 
consider two representative cases and show the 
time-evolution of $\theta$ (in units of $\pi$) over $20$ 
cycles of the field.  
In each case, the parameters considered lie deep in the respective phases 
in $F$-$\Omega$ space 
and the dynamics quickly settle the bar into the appropriate alignment.  
The difference in the distribution of the SP in the two cases is striking. 
In the regular (vertical 
orientation) case the particles are essentially confined to the regions along 
the perimeter of the 
cell, and especially in the corners. This is in contrast to the 
anomalous 
case where the particles inhabit a much larger fraction of the box and 
are 
only excluded from two narrow regions immediately surrounding
the (large) charges on the bar. Thus, in the 
high-frequency case the rapid motion of the bar has a cavitating effect, 
which leads the SP to generate a very flat-bottomed effective potential 
well; hence, the orientation of the bar is principally determined by the 
external field as if in a vacuum.  By contrast, in the low-frequency case 
the cloud of SP fills a larger space, contracting
and expanding in synchrony with the bar's oscillations.  As a result, the 
vertical gradient in the cloud's density produces a net torque on the bar, 
and the energetically favored configuration is the one in which the PP is 
kept horizontal by the SP's effective potential, which prevails on the 
external potential in orienting of the rod.				       
The fact that the SP are clustered closer to the bar 
for anomalous orientation provides insight into the relative energetics of 
the two configurations. The decreased mean spacing
between each of the SP and the bar means that
this configuration is a more energetic and, in a dynamical sense, unstable one.
Finally, in the random phase neither the external field nor the pressure
of the cloud is dominant.  It is possible that this regime is
governed by single SP proximity events rather than collective behavior. 
\begin{figure}
\includegraphics[height=3.1in,width=3.1in]{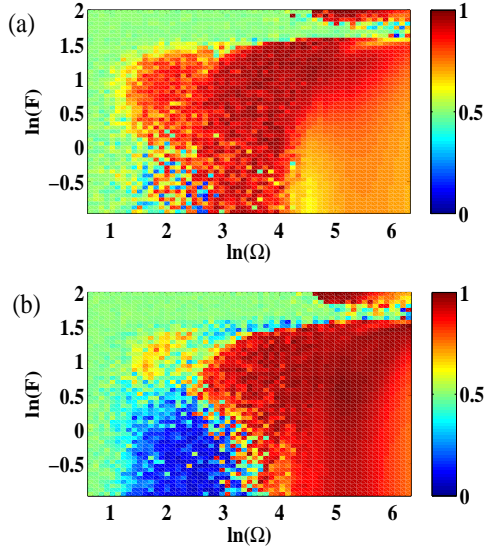}
\caption{\label{fig3} Contrasting cases of (a) rectangular and (b) square aspect ratio.
$R$, $V$, and $H$ denote the random, vertical and horizontal phases respectively.}
\end{figure}

The simulations shown so far have been for a rectangular ($3:5$) box, so 
one wonders how the phase diagram in $F$-$\Omega$ changes when the aspect ratio 
is square ($4:4$). As seen from Fig.~(\ref{fig3}), the anomalous phase 
vanishes, while the phase boundaries are still visible. This
suggests that the distribution of SP no longer generates
adequate screening for the anomalous orientation to be attained. The question 
of why the rectangular case should be reflective of what happens in 
the experiments is obviously
beyond the scope of our simplified single-bar model. We speculate that, in laboratory systems,
the interactions among multiple PP in solution favor a mutual alignment of
the bars in a lattice structure with spacing consistent with a rectangular cell.  We further
speculate that such a putative statistical bias towards "staggering" of the bars may depend
only weakly on the concentration of the PP; in fact, the experiments \cite{cecco} show that PP 
concentration does not affect the anomalous orientation significantly.  These conjectures
will be the object of future investigations.
Finally, as suggested earlier, the sharp boundaries between the orientation regions in $F$-$\Omega$ 
space suggest a three-phase diagram akin to what has been seen, for example, in disordered 
spin systems~\cite{spinref}. There are two distinct ordered phases, corresponding to the 
normal and anomalous orientation of the rod-like colloid, and a disordered or ``glassy" phase 
corresponding
to random orientation. Our dynamical model makes clear the competing influences (frustration) 
inherent in the system, and this analogy may prove useful in explaining features like 
re-entrancy visible in the phase diagrams.


\begin{thebibliography}{}
\bibitem{saville} W.B. Russell, D.A. Saville and W.R. Schowalter,
{\it Colloidal dispersions} (Cambridge University Press, Cambridge 1989).
\bibitem{hunter} R.J. Hunter, {\it Foundations of colloid science},
(Clarendon Press, Oxford 1987).
\bibitem{kramer} H. Kramer, M. Deggelman, C. Graf, M. Hagenbuchle, C. Johner and R. Weber,
Macromolecules {\bf 25}, 4325 (1992).
\bibitem{cates} M. E. Cates, J. Phys. II {\bf 2}, 1109 (1992).
\bibitem{blair} M. J. Blair and G. N. Patey, J. Chem Phys. {\bf 111}, 3278 (1999).
\bibitem{melissa} M. Rotunno, T. Bellini, Y. Lansac, and M.A. Glaser,
J. Chem Phys. {\bf 121}, 5541 (2004).
\bibitem{cecco} F. Mantegazza, M. Caggioni, M.L. Jimenez and T. Bellini,
Nature Physics {\bf 1}, 103 (2005).
\bibitem{cecco2} T. Bellini, F. Mantegazza, V. Degiorgio, R. Avallone and
D.A. Saville, Phys. Rev. Lett. {\bf 82}, 5160 (1999).
\bibitem{spinref} D. Chowdhury, {\it Spin glasses and other frustrated systems}, (Princeton University Press, Princeton 1986).
\end{thebibliography}
\end{document}